\documentclass[twocolumn,pra,aps]{revtex4-1}
\usepackage{amsmath}
\usepackage{amsfonts}
\usepackage{amssymb}
\usepackage{graphicx}
\usepackage{hyperref}
\usepackage{bm}
\usepackage{color}
\usepackage{footmisc}

\usepackage{times}
\usepackage{dsfont}
\usepackage[squaren,Gray]{SIunits}

\hypersetup{backref,
colorlinks=true,
linkcolor=blue,
linktoc=black,
citecolor=cyan,
urlcolor=black}

\newcommand{\pc}[1]{\ensuremath{\left(#1\right)}} 
\newcommand{\px}[1]{\ensuremath{\left\lbrace#1\right\rbrace}} 
 
\newcommand{\ket}[1]{\ensuremath{\left\vert#1\right\rangle}} 
\newcommand{\md}[1]{\ensuremath{\left\vert#1\right\vert}}

\begin{document}
\title{
Anisotropic light-shift and magic-polarization\\ of the intercombination line of Dysprosium atoms in a far-detuned dipole trap
}

\author{Thomas Chalopin$^{1}$}
\author{Vasiliy Makhalov$^{1}$}
\author{Chayma Bouazza$^{1}$}
\author{Alexandre Evrard$^{1}$}
\author{Adam Barker$^{2}$}
\author{Maxence Lepers$^{3,4}$, Jean-Fran{\c c}ois Wyart$^{3,5}$, Olivier Dulieu$^{3}$}
\author{Jean Dalibard$^{1}$}
\author{Raphael Lopes$^{1}$}
\email{raphael.lopes@lkb.ens.fr}
\author{Sylvain Nascimbene$^{1}$}
\address{${}^{1}$Laboratoire Kastler Brossel, Coll\`ege de France, CNRS, ENS-PSL Research University, UPMC-Sorbonne
Universit\'es, 11 place Marcelin Berthelot, F-75005 Paris, France}
\address{${}^{2}$Clarendon Laboratory, University of Oxford, Oxford OX1 3PU, United Kingdom}
\address{${}^{3}$Laboratoire Aim{\'e} Cotton, CNRS, Universit{\'e} Paris-Sud, ENS Paris-Saclay, Universit{\'e} Paris-Saclay, 91405 Orsay, France}
\address{${}^{4}$Laboratoire Interdisciplinaire Carnot de Bourgogne, CNRS, Universit\'e de Bourgogne Franche-Comt\'e, 21078 Dijon, France}
\address{${}^{5}$LERMA, Observatoire de Paris-Meudon, PSL Research University, Sorbonne Universit{\'e}s, UPMC Univ.~Paris 6, CNRS UMR8112, 92195 Meudon, France}

\begin{abstract}
We characterize the anisotropic differential ac-Stark shift for the Dy $626$~nm intercombination transition, induced in a far-detuned $1070$~nm optical dipole trap, and observe the existence of a ``magic polarization" for which the polarizabilities of the ground and excited states are equal.
From our measurements we extract both the scalar and tensorial components of the dynamic dipole polarizability for the excited state, $\alpha_E^\text{s} = 188 (12)\,\alpha_\text{0}$ and $\alpha_E^\text{t} = 34 (12)\,\alpha_\text{0}$, respectively, where $\alpha_\text{0}$ is the atomic unit for the electric polarizability. We also provide a theoretical model allowing us to predict the excited state polarizability and find qualitative agreement with our observations. Furthermore, we utilize our findings to optimize the efficiency of Doppler cooling of a trapped gas, by controlling the sign and magnitude of the inhomogeneous broadening of the optical transition. The resulting initial gain of the  collisional rate allows us, after forced evaporation cooling, to produce a quasi-pure Bose--Einstein condensate of $^{162}$Dy with $3\times 10^4$ atoms.
\end{abstract}
\maketitle

Lanthanide atoms offer a new and exciting test bed on which to explore long-awaited physical phenomena such as the appearance of the roton-excitation in dipolar Bose--Einstein condensates, due to their large magnetic moments, \cite{Santos:2003,Wilson:2008,Boudjemaa:2013,Chomaz:2018} or the occurrence of exotic superfluid phases based on narrow transition lines and a dense Feshbach resonance spectrum \cite{Nascimbene:2013,Cui:2013,Burdick:2016,Ferrier-Barbut:2016,Kadau:2016}.

These unique properties arise thanks to the partially-filled, submerged 4f shell but, due to the large number of unpaired electrons, come with a drawback in terms of complexity. For instance, the dynamic (dipole) polarizability, which is of fundamental importance as it sets the strength of light-matter interactions, is theoretically challenging to estimate \cite{Dzuba:2011, Li:2017a}. Several experimental efforts have been made to benchmark these theoretical models but have, so far, mainly addressed the polarizability of the ground state \cite{Kao:2017, Ravensbergen:2018, Becher:2018}.
 
In the case of the $626 $~nm intercombination transition used in several dysprosium (Dy) cold atom experiments, little is known about the excited state polarizability \cite{Maier:2014, Lucioni:2017,Lucioni:2018}. 
Besides its fundamental interest, its characterization plays an important role when considering the action of near-resonant light on a gas confined in the high-intensity field of an optical dipole trap \cite{Lundblad:2010}. 
In particular, when the ground and excited states have different polarizabilities, one expects a differential light-shift in the resonance line proportional to the trapping light intensity.

If the differential light-shift is close to or larger than the linewidth of the transition, the light-matter interaction becomes strongly affected by the trapping optical beam. In particular, due to the spatial variation of the light intensity, the coupling becomes spatially-dependent. This effect received much attention in the case of atomic clocks since it couples the external and
internal degrees of freedom, degrading the coherence of spectroscopic measurements. For alkali \cite{Lundblad:2010} and alkaline-earth atoms, the existence of ``magic-wavelengths'' helped to suppress this nuisance \cite{Ido:2000, Katori:2003,Takamoto:2005, Ye:2008,Katori:2009,Ludlow:2015}.
Furthermore, the line-shift induced by the presence of off-resonant optical traps also affects the laser cooling efficiency \cite{Ido:2003, Chalony:2011} and can be used to spatially tailor light-matter interactions \cite{Stellmer:2013}.

For lanthanide atoms, due to the significant tensorial contribution to the total atomic polarizability, the differential light-shift strongly depends on the trapping light polarization \cite{Kao:2017, Becher:2018}.
This offers the possibility to locally vary the transition resonance frequency by fine-tuning the trapping beam polarization; this feature has also been applied in a similar manner to alkali atoms, using the differential vectorial polarizability \cite{Kim:2013}. The magic-wavelength behaviour is then replaced by a ``magic-polarization''.

\begin{figure}[t!]
\centering 
  \includegraphics[width=\columnwidth]{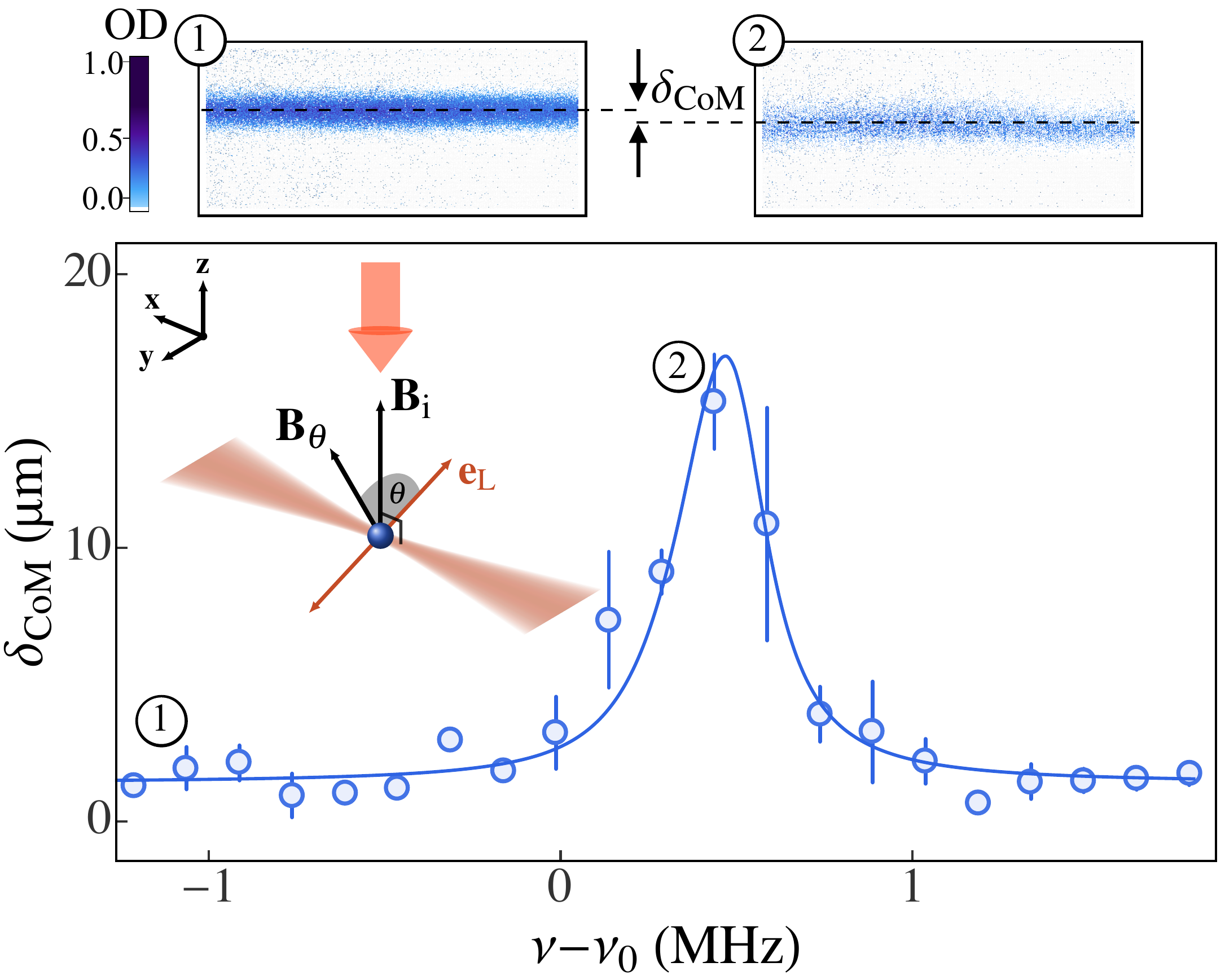}
\caption{\label{Fig1} Center-of-mass displacement resonance. Schematic drawing: a near-resonant ($626$~nm) beam is applied to a cold atomic sample optically trapped around the focal point of a 1070~nm laser beam propagating along the $x$ axis. The magnetic field bias is orientated in a plane perpendicular to the optical beam propagation axis, forming an angle $\theta$ with the polarization vector $\mathbf{e}_\text{L}$. Two orientations of $\mathbf{B}$ are represented: the initial vertical orientation ($\mathbf{B}_\text{i}$) and the value corresponding to the resonance curve shown in the main panel ($\mathbf{B}_\theta$). The beam is applied for a short duration and accelerates the atoms, leading to a displacement of the cloud center-of-mass (CoM), measured after time-of-flight (ToF) represented by the dashed lines (see top panels). The center-of-mass displacement ($\delta_\text{CoM}$) is plotted as a function of the laser frequency $\nu$ for $\theta = 80^\circ$ and fitted using Eq.~\eqref{eq:com} with the free parameter $\Delta \alpha (\theta)$. The error bars denote the r.m.s. deviation of 3 independent measurements.}
 \end{figure}

In this Article we characterize the anisotropic differential light-shift in the case of the Dy $626$~nm transition ($\ket{g}=\ket{J = 8,m_\mathit{J}=-8} \rightarrow \ket{e}=\ket{J'=9,m_\mathit{J'}=-9}$) for a cold gas trapped in a far-detuned $1070$~nm optical trap (see Fig.~\ref{Fig1})\footnote{The $1070$~nm laser light is provided by a commercial multi-mode IPG (50W) laser. A similar magic-polarization behaviour was observed using a $1064$~nm single-mode Azur Light (45W) laser.}. Using theoretical predictions for the polarizability of the ground state \cite{Dzuba:2011,Li:2017} (see also measurements of Ref.~\cite{Ravensbergen:2018}), we extract the excited state polarizability, and identify a tensorial component of much larger amplitude than for the ground state. By tuning the relative angle between the laser polarization and an external magnetic field, we find a magic-polarization for which the differential light-shift between $\ket{g}$ and $\ket{e}$ is cancelled. We compare our results to a theoretical model described in Section~\ref{sec:theory} and find qualitative agreement.
 As a concrete example of the relevance of this magic-polarization behaviour, we implement a one-dimensional Doppler cooling experiment which we optimize by adjusting the spatially-dependent differential light-shift. We observe a significant gain in the collisional rate for the case of a small, positive differential light-shift which leads to an enhanced $(\text{red})$ detuning of the cooling light at the trap center . We interpret this result as a suppression of light-assisted collisions at the bottom of the potential where the atomic density is higher, while cooling remains efficient in the wings. This cooling stage allows us to significantly boost the cloud initial phase-space density, and, after a $4$~s forced evaporation procedure, to reach quantum degeneracy for a cloud of $^{162}$Dy at a critical temperature $T_c \approx 120 (20) $~nK and atom number $N \approx 7 \times 10^4$.

\section{Differential light-shift: Magic polarization}
\label{sec:DLS}

The interaction between an atom and a monochromatic laser field of frequency $\omega$ gives rise to two types of effects. First, the atom can scatter photons into empty modes of the electromagnetic field via spontaneous emission processes. Second, each atomic level may be shifted by the light field (AC-Stark shift). Here we restrict ourselves to the case of a non-resonant light field, which in our case corresponds to the laser beam used for trapping the atoms, so that the first type of effect is negligible and we focus on the latter. 

Let us consider, for instance, the atomic ground level $G$, with angular momentum $J$. At lowest order in laser intensity, the atom-light interaction leads to stimulated Raman processes in which the atom passes from the Zeeman state $|G,J,m\rangle$ to another state  $|G,J,m'\rangle$ with $|m'-m|\leq 2$. The light-shift operator is then a rank 2 tensor acting on the  manifold $G$. It can be expressed in terms of the dynamic polarizability, $\underline{\underline{\mathcal{\alpha}}}_G(\omega)$, with scalar ($\alpha^\text{s}_G$), vectorial ($\alpha^\text{{v}}_G$) and tensorial ($\alpha^\text{t}_G$) contributions.

For a  laser beam with linear polarization $\mathbf{e}_\text{L}$, the vectorial contribution is suppressed by symmetry and the restriction of the atom-light interaction to $G$ can be written 
\begin{align}
\hat H_\text{a-l,G} &= \tilde{V}(\mathbf{r}) \px{ \alpha_G^\text{s}  \mathds{\hat 1} +\alpha_G^\text{t} \frac{3 \pc{ \mathbf{\hat J}\cdot \mathbf{e}_\text{L}}^2 -  \mathbf{\hat J}^2}{J(2J-1)} },
\end{align}
where $\mathbf{\hat J}$ is the angular momentum operator. Here $\tilde{V}(\mathbf{r}) = -\frac{1}{2 \epsilon_\text{0} c} I(\mathbf{r}) $ where $I(\mathbf{r})$ is the  laser beam intensity, $\epsilon_\text{0}$ the vacuum permittivity and $c$ the speed of light.

In the presence of a static magnetic field $\mathbf{B}$, the Hamiltonian describing the dynamics within $G$ is thus:
\begin{align}
\label{eq:fullH}
\hat H_G =\hat  H_\text{0,G} + \hat H_\text{a-l,G}\,, 
\end{align}
with\hspace{0.9em} $\hat H_\text{0,G}= g_J \mu_\text{B} \, \mathbf{J}\cdot \mathbf{B}$, $g_J$ the Land\'e $g$-factor and $\mu_\text{B}$ the Bohr magneton. Let us assume for now that the tensorial contribution to $\hat H_\text{a-l,G}$ can be treated at first-order in perturbation theory with respect to $\hat H_\text{0}$ (this assumption will be released later). The energy shift for the state of lowest energy $\ket{g}$ in the manifold $G$ is then given by 
\begin{align}
E_g= E_{g,\text{0}} + \tilde{V}(\mathbf{r}) \px{\alpha_G^\text{s} + \frac{\alpha_G^\text{t}}{2} \pc{3 \cos^2 \theta -1}},
\end{align}
where  $\theta$ is the angle between the static magnetic field and the beam polarization (see Fig.\ref{Fig1}). 

A similar analysis can be performed for any relevant excited electronic level $E$, in particular the one used here for Doppler cooling. The energy difference between the states of lowest energy in each manifold, $\ket{g} $ and $ \ket{e}$, is  equal to
\begin{align}
h \nu_\text{0}' (\mathbf{r}) = h \pc{\nu_\text{0}  +  \Delta \nu_\alpha  (\mathbf{r})} ,
\label{eq:resonanceVsV}
\end{align}
where  $h \nu_\text{0} = \Delta E_\text{0} = E_{e,\text{0}} - E_{g, \text{0}} $, and
\begin{align}
h \Delta \nu_\alpha (\mathbf{r}) = \tilde{V}(\mathbf{r}) \Delta \alpha \, ,
\label{eq:deltaalpha}
\end{align}
with $\Delta \alpha = \Delta \alpha^\text{s} +\frac{1}{2}\Delta \alpha^\text{t} \pc{ 3 \cos^2 \theta -1 }$ and  $\Delta \alpha^{\text{(s, t)}} = \alpha_{E}^{\text{(s, t)}} - \alpha_{G}^{\text{(s, t)}} $
Importantly, for $\md{\Delta \alpha^\text{t}/\Delta \alpha^\text{s} -1/2} \geq 3/2$, the differential light-shift cancels for a specific polarization angle $ \theta_\text{magic} = \arccos\pc{\sqrt{\frac{1}{3} \pc{1 - 2\frac{\Delta \alpha^\text{s}}{\Delta \alpha^\text{t}}}}}$, that we will refer to as a magic-polarization angle in the following text.

We begin by producing a cold sample of $10^7$ $^{164}$Dy atoms in the state $\ket{g}$, held in a 1070~nm dipole trap beam.
The beam polarization is linear and oriented at approximately $ 60^\circ$ relative to the magnetic field ($\mathbf{B}_\text{i}$) initially aligned with the vertical $\mathbf{\hat{z}}$ axis (see Fig.~\ref{Fig1}); as shown hereafter this seemingly arbitrary angle corresponds to $\theta_\text{magic}$. The magnetic field is then re-orientated to probe different values of $\theta$. The duration of the re-orientation is chosen long enough for the atoms to follow adiabatically the state $\ket{g}$ \footnote{As discussed in Sec.~\ref{sec:theory}, for the ground level polarizability the scalar component is much larger than the tensorial one. Consequently, the rotation of the magnetic field does not modify the trapping potential experienced by the atoms in $\ket{g}$.}.

In order to probe the resonance frequency for the $\ket{g} \rightarrow \ket{e}$ transition, we apply for $\tau = 30$~\micro s a near-resonant beam, circularly polarized ($\sigma^-$) and propagating along $\mathbf{\hat{z}}$ \footnote{We use a collimated beam with a waist of $5.5$~mm, much wider than the cloud spatial extension and therefore uniform for the sample. We have also performed several experiments with different pulse durations but did not observe significant changes of the differential light-shift for pulse lengths varying between $10$ up to $100$\micro s . The $30$~\micro s pulse length thus corresponds to the shortest pulse guaranteeing a good signal-to-noise ratio.}.
In the limit of a short pulse, the momentum kick experienced by the atoms reaches its maximum value when the laser frequency equals the transition frequency. This leads to a maximum displacement of the cloud center-of-mass (CoM) after time-of-flight (ToF), allowing us to extract, as a function of the dipole trap intensity, the transition resonance frequency and the differential light-shift.

  \begin{figure}[t!]
\centering 
  \includegraphics[width=\columnwidth]{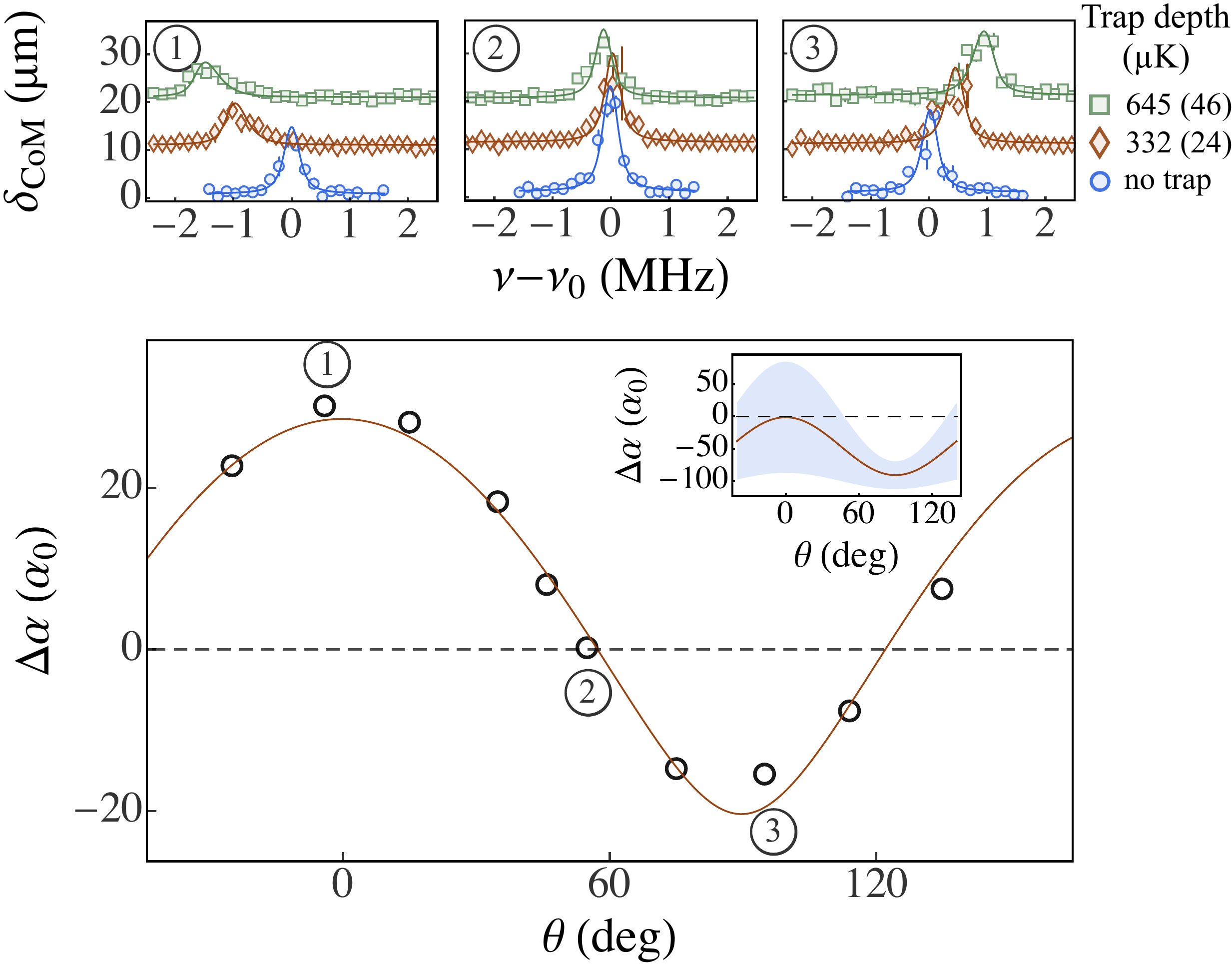}
\caption{\label{Fig2} Differential light-shift as a function of the relative angle $\theta$. Top panels: CoM resonances as a function of the trap depth experienced in $\ket{g}$ (see legend) for three different angles : $0^\circ ,\, 55^\circ \, \text{and} \, 100^\circ$. The CoM values have been shifted with respect to each other for clarity. The error bars denote the r.m.s. deviation of 3 independent measurements.
Main panel: $\Delta \alpha$ as a function of $\theta$. The solid line corresponds to a fit based on the energy difference between excited and ground states following the diagonalization of $\hat{H}$ given in Eq.~\eqref{eq:fullH} with $\Delta \alpha^\text{s}$ and $\Delta \alpha^\text{t}$ as free parameters. Inset: Differential polarizability as a function of $\theta$ using Eq.~\eqref{eq:deltaalpha} and the theoretical values given in Section~\ref{sec:theory}. The shaded region represents the differential polarizability uncertainty.}
 \end{figure} 

In more detail, the mean radiative force exerted on an atom at a position $\mathbf{r}$ is given by  \cite{Foot:2005}
\begin{align}
\mathbf{F}(\Delta \omega(\mathbf{r}),v_z)=-\hbar k \frac{\Gamma}{2} \frac{s_\text{0}}{1+s_\text{0}+ 4 \pc{\frac{ \Delta \omega(\mathbf{r})-k v_z}{\Gamma}}^2}\, \mathbf{\hat{z}}\, ,
\end{align}
where $k = 2\pi / \lambda$ is the recoil momentum, with $\lambda = 626 \, \text{nm}$, $s_\text{0} = I_\text{0}/I_\text{sat}$ the saturation parameter with $I_\text{sat} = 72  \, \text{\micro W} /\text{cm}^2$, $\Delta \omega (\mathbf{r})  = 2\pi \times \pc{\nu - \nu_\text{0}' (\mathbf{r})} $, $v_z$ the atomic velocity along $\mathbf{\hat{z}}$, and $\Gamma = 2\pi \times 136$ kHz the transition linewidth. During the application of this pulse the cloud displacement is negligible (on the order of $1$-$2$~\micro m) and the only sizable effect of the pulse is a sudden change of the atomic velocity. Furthermore, the acquired Doppler shift during the pulse is negligible compared to $\Gamma$.
The optical dipole trap is then switched off and an absorption image is taken after a ballistic expansion of duration $t_\text{ToF} =1.5$~ms. The momentum kick as a result of the pulse translates into a center-of-mass (CoM) position shift, $\delta_\text{CoM}$, given by
\begin{align}
\delta_{\text{CoM}} =\frac{t_{\text{ToF}}}{m} \tau \int \text{d} \mathbf{v}\,  \text{d}\mathbf{r} \;\; n(\mathbf{r},\, \mathbf{v}) \, F \pc{\Delta \omega (\mathbf{r},\mathbf{v}) },
\label{eq:com}
\end{align}
where $m$ is the atom mass and $n(\mathbf{r},\, \mathbf{v})$ is the normalized spatial and velocity distribution of the cloud, computed for an initial cloud temperature $T \approx 100$~\micro K and a harmonic trapping potential with frequencies $\px{\omega_x, \omega_{y, z}} = 2\pi \times \px{9(1)\, \text{Hz},1.9(1)\, \text{kHz} }$.

In Fig.~\ref{Fig1} we show a typical CoM-displacement resonance as a function of the laser frequency, $\nu$. The origin of the frequency axis is set by the bare resonance frequency, $\nu_\text{0}$, that we extract from a similar resonance measurement performed in the absence of the trapping beam \footnote{We checked that neither the red-pulse duration length ($10<\tau<100$~\micro s) nor the cloud temperature affected substantially the value of $\nu_\text{0}$ \label{footnote1}}. 
Using Eq.~\eqref{eq:com} we record, for different values of $\tilde{V}(\mathbf{0})$ the resonance position $\nu_\text{0}'$ (see Fig.~\ref{Fig2} top panels). We verify that $ \nu_\text{0}'$ varies linearly with $\tilde{V}(\mathbf{0})$ and extract $\Delta \alpha (\theta)$ from the slope. The same procedure is then repeated for several orientations of the magnetic field $\mathbf{B}_\theta$ thus probing different relative angles $\theta$ (see Fig.~\ref{Fig1}). 

We recover the expected dependence of the total polarizability difference, $\Delta \alpha$, as a function of $\theta$, as shown in Fig.~\ref{Fig2}~(main panel). We observe that $\Delta \alpha = 0$ for $\theta_\text{magic}~=~57(2)^\circ $, corresponding to a cancellation of the differential light-shift, and characteristic of magic-polarization behaviour. 
The fitting function shown in Fig.~\ref{Fig2} (main panel) corresponds to the differential light-shift computed numerically from the energy difference between the state of lowest energy ($\ket{g}$) of Eq.~\eqref{eq:fullH} and its equivalent solution for the excited state manifold ($\ket{e}$),
with free parameters $\Delta \alpha^\text{s}$ and $\Delta \alpha^\text{t}$. We find $\Delta \alpha^\text{s} = -5(2)\, \alpha_\text{0}$ and $\Delta \alpha^\text{t} = 33 (2)\, \alpha_\text{0}$,  where $\alpha_\text{0} =4\pi \epsilon_\text{0} a_\text{0}^3 $ and $a_\text{0}$ is the Bohr radius. Using the theoretical values of $\alpha_G^{\text{(s, t)}}$ (see Section~\ref{sec:theory}) we determine the excited state scalar and tensorial polarizabilities $\alpha_{E}^{\text{s}} = 188 (12)\, \alpha_\text{0}$ and $\alpha_{E}^{\text{t}} = 34(12) \,\alpha_\text{0}$, respectively. The small error bars reported here are purely statistical but systematic effects can play an important role in the quantitative determination of $ \alpha_{E}^{\text{(s, t)}} $. For instance, deviations from the theoretical values of $\alpha_G^\text{(s, t)}$, such as the ones reported for 1064~nm (see Ref.~\cite{Ravensbergen:2018}), would automatically shift the reported absolute values of $\alpha_{E}^\text{(s, t)}$. However, the existence of the magic polarization angle ($\theta_\text{magic}$) is robust with respect to these systematic effects. 

Our observations imply that, although the scalar components of the dynamic polarizability are similar for both states, the tensorial contribution of the excited state is much larger than for the ground state. Note however that the tensorial component of the excited state does not alone fulfill the condition  $\alpha^\text{t}_E  > 2 \alpha^\text{s}_E$ needed to cancel the light-shift of that state.

\section{Theoretical estimation of the excited state polarizability}
\label{sec:theory}

The scalar $\alpha^\text{s}$ and tensor polarizabilities $\alpha^\text{t}$ are calculated using the sum-over-state formula (see \textit{e.g.}~\cite{Li:2017}). For the ground state, the data of Ref.~\cite{Li:2017} give $\alpha^\text{s}_G = 193\, (10) \, \alpha_0$ and $\alpha^\text{t}_G = 1.3 (10) \, \alpha_0$ at 1070~nm \footnote{To be noted that the data of Ref.~\cite{Li:2017} slightly differs from the recent calibration of the ground state polarizability in the presence of a 1064~nm optical beam reported in Ref.~\cite{Ravensbergen:2018} in which $\alpha_G^\text{s} = 184.4 \, (2.4) \; \alpha_0$ and $\alpha_G^\text{t} = 1.7\, (6) \; \alpha_0$ were found.}.

For the excited state $\ket{e}$ considered above, the energies and transition dipole moments (TDMs) towards even-parity levels are required to estimate the polarizability.
For levels belonging to configurations that were observed experimentally, energies and TDMs were explicitly calculated with the semi-empirical method implemented in Ref.~\cite{Cowan:1981}, which has been extended by some of us \cite{Li:2017, Li:2017a, Lepers:2018}. Those levels are split into three groups of configurations: (i) $4f^{10}6s^2 + 4f^{10}5d6s + 4f^96s^26p$; (ii) $4f^{10}6s7s + 4f^{10}6s6d$, and (iii) $4f^95d6s6p$ \footnote{The xenon-core configuration [Xe] preceding the configurations has been omitted for clarity.}. Following Ref.~\cite{Lepers:2018}, we multiply the relevant monoelectronic TDMs by a scaling factor (0.794 for $\langle ns|\hat{r}|n'p\rangle$, 0.923 for $\langle nf|\hat{r}|n'd\rangle$ and 0.80 for $\langle nd|\hat{r}|n'p\rangle$), in order to improve the least-square fit of the measured TDMs by the calculated ones.
Some unobserved levels are likely to significantly contribute to the polarizability; for instance, those belonging to the $4f^{10}6p^2$ configuration. We account for those levels using the effective model of Ref.~\cite{Li:2017a}, with configurations $4f^{10}6p^2$, $4f^{10}6sns$ ($n=8$ to 10) and $4f^{10}6snd$ ($n=7$ to 9). Transition energies are calculated using the corresponding observed energy levels in ytterbium, while monoelectronic TDMs are the \textit{ab initio} values multiplied by the scaling factors given above.
Overall we find $\alpha^\text{s}_E = 132 \,(33) \, \alpha_0$ and $\alpha^\text{t}_E = 61 \,(33) \, \alpha_0$. 

As shown in Fig.~\ref{Fig2} (inset) our model is consistent, within error bars, with the experimental observation of a magic-polarization. Such agreement relies on a large difference between the tensorial contributions of the excited and ground states.
The predicted magic polarization angle ($-60^\circ<\theta <60^\circ $) although in qualitative agreement with our observations does not allow one to quantitatively account for our results. This is due to the aforementioned difficulty to accurately resolve the excited state polarizability which leads to a large differential polarizability uncertainty.

\section{Application to Doppler cooling}

We demonstrate the relevance of a magic-polarization by considering Doppler cooling in an optical dipole trap \cite{Lett:1989,Dalibard:1989,Schmidt:2003,Maier:2015th, Chalony:2011}. This process is implemented in order to significantly reduce the cloud temperature over a short timescale, typically set by the weakest trapping frequency.
For this purpose we use the $626$~nm transition considered above where $\Gamma = 2\pi \times 136$~kHz. Since $\Gamma$ is small compared to the typical differential light-shifts reported in Fig.\ref{Fig2}, one expects the cooling efficiency to be strongly dependent on the optical beam polarization.

  \begin{figure}[t!]
\centering 
  \includegraphics[width=\columnwidth]{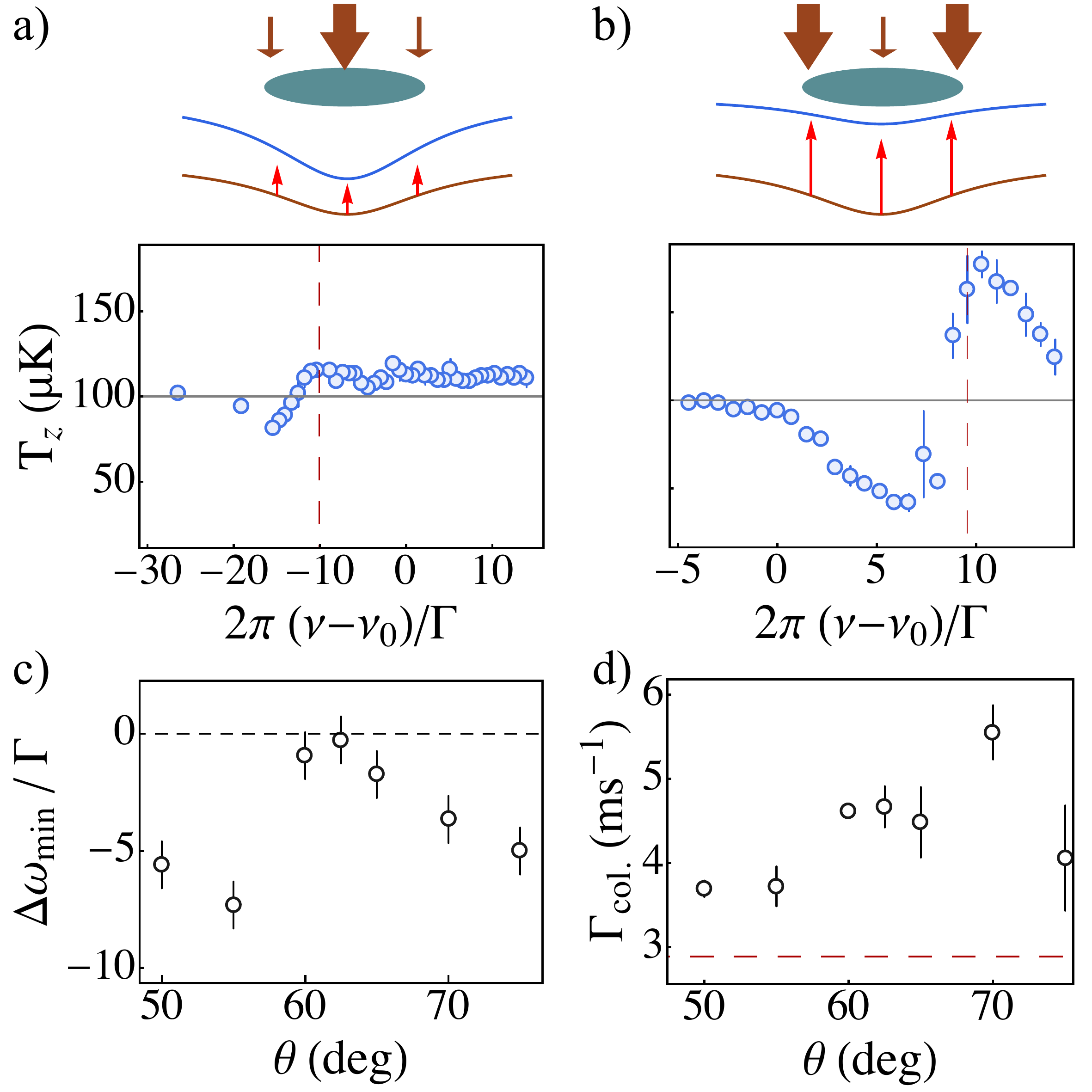}
\caption{\label{Fig3} Doppler cooling efficiency as a function of $\theta$ and gain in phase-space density: 
Cooling efficiency for a) $\theta = 50^\circ$ and b) $\theta = 75^\circ$  as a function of the cooling beam frequency $\nu$, for $s_0 = 0.5$ and a pulse time $\tau = 20$~ms. The vertical red dashed line indicates the transition resonance at the trap center.
c) Detuning from the trap center for which the minimal temperature is recorded ($\Delta \omega_\text{min}$) as a function of $\theta$. The black dashed line indicates the zero-detuning limit.
d) Collisional rate, $\Gamma_\text{col.}$, as a function of $\theta$ . An optimum is visible for $\theta = 70^\circ$ corresponding to a small, positive differential light-shift. The horizontal dashed red line corresponds to the value of $\Gamma_\text{col.}$ prior to the  Doppler cooling stage.
The error bars denote the r.m.s. deviation of 3 independent measurements.
}
 \end{figure} 

In order to optimize the cooling efficiency we vary slightly the value of $\theta$ around the magic polarization angle $\theta_\text{magic}$ (see Fig.~\ref{Fig3})\footnote{A $2$~G bias field ensures that the Zeeman splitting is much larger than the transition linewidth. Therefore, small imperfections of the laser beam polarization are not particularly relevant. Furthermore, the weak Clebsch-Gordan coefficients for $\pi$ and $\sigma^{+}$ transitions compared to the $\sigma^{-}$  transition render these imperfections even less relevant.}. We observe two regimes with distinct behaviour. In the case of a negative differential light-shift ($\Delta \nu_\alpha (\mathbf{r})<0$), the cooling is inefficient. 
On the other hand, for small, positive values of the differential light-shift, the cooling stage is efficient and leads to an increased collisional rate ($\Gamma_\text{col.}$). The qualitative explanation for that behaviour is summarized schematically in Fig.~\ref{Fig3} (top panels). In the first case, the denser, central, region of the atomic cloud is, due to the strong negative differential light-shift, closer to resonance and therefore interacts strongly with the cooling beam. However the local density is large and light-assisted collisions are predominant; this results in a very poor cooling efficiency as shown in Fig.~\ref{Fig3}a. For the case of a positive differential light-shift the situation is reversed. The central region is strongly detuned, and light-assisted collisions are reduced while cooling taking place in the wings, where the density is lower, is very efficient (see Fig.~\ref{Fig3}b).

  \begin{figure}[t!]
\centering 
  \includegraphics[width=\columnwidth]{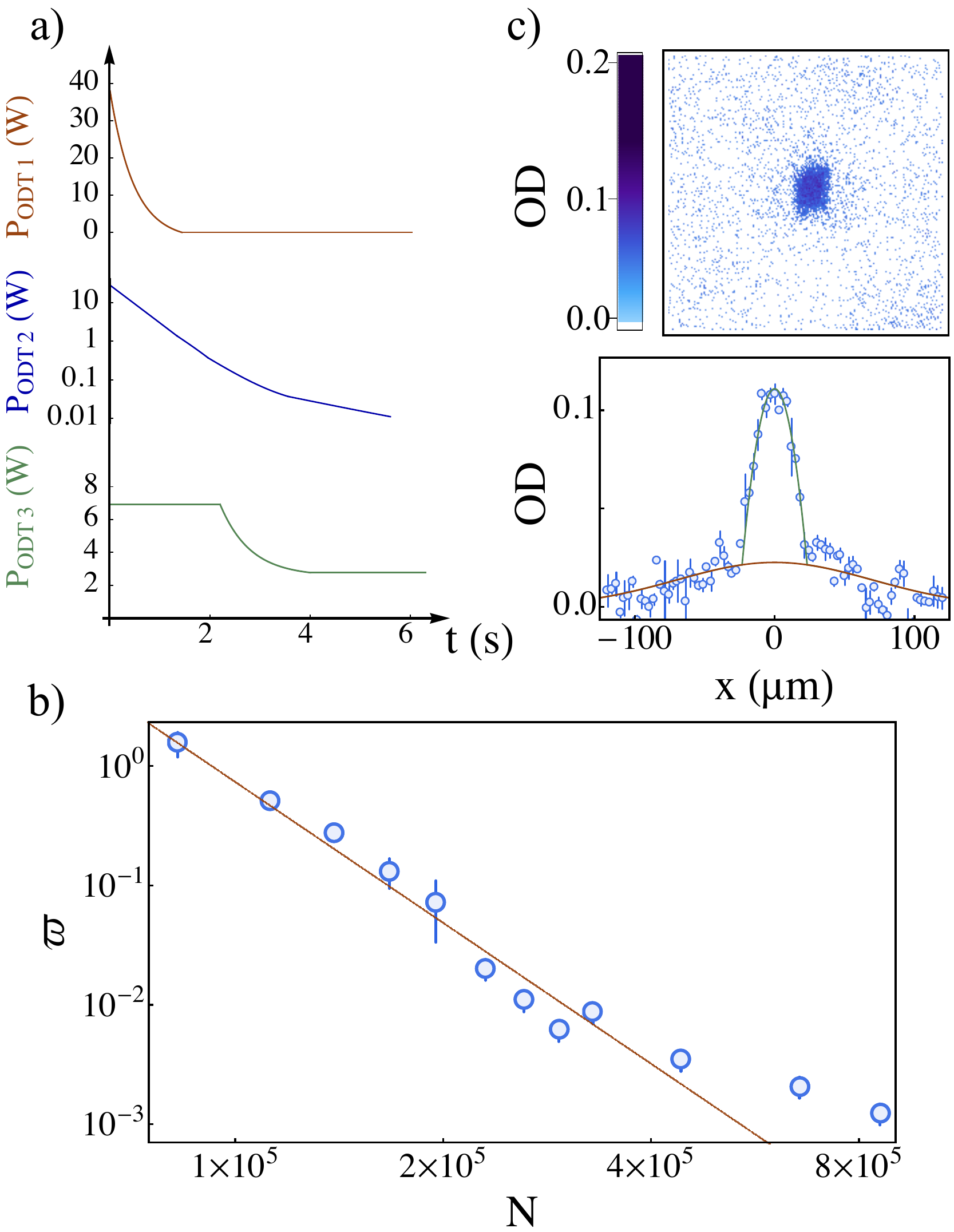}
\caption{\label{Fig4}Condensation of $^{162}$Dy: 
a) Schematic representation of the evaporation procedure in the optical dipole trap. 
b) Phase-space density, $\varpi$, as a function of the atom number $N$ in logarithmic scale.
c) Two-dimensional picture and integrated profile of a dipolar Bose-Einstein condensate with a condensed fraction of 50$\%$. The solid line corresponds to a gaussian plus parabolic fit.
}
 \end{figure} 

To better understand the above, empirical description of the cooling and heating mechanisms at work, we also report the detuning frequency at which the minimal temperature is recorded for several different values of $\theta$ (see Fig.~\ref{Fig3}c). The detuning is expressed with respect to the resonance frequency at the trap center, such that $\Delta \omega_\text{min} = 2\pi \times (\nu_\text{min} - \nu_0'(\mathbf{0}))$. In the case of a negative differential light-shift we observe an optimum cooling efficiency for large negative detuning values. This behaviour suggests that the cooling beam is also responsible for local heating and losses at the trap center; processes which are minimized by increasing the absolute frequency detuning. In the case of a differential light-shift cancellation, the detuning is compatible with the textbook $-\Gamma/2$ result. For positive differential light-shifts we also observe an optimum at	 an enhanced negative detuning. This is expected since the cooling mechanism mainly occurs in the outer regions of the cloud, where the differential light-shift is smaller and therefore the frequency detuning from the trap center is larger (see Fig~\ref{Fig3} top panels).

We optimize the cooling efficiency by maximizing the collisional rate $\Gamma_\text{col.}$, which is a natural figure of merit towards achieving Bose-Einstein condensation. For each value of $\theta$ we maximize $\Gamma_\text{col.}$ by adjusting the frequency and $\tau$ for a fixed $s_0 = 0.5 $. As shown in Fig.~\ref{Fig3}d, we observe that for small positive differential light-shifts ($\theta \approx 70^\circ$) a maximum is reached. A similar method has also been applied to reach the quantum limit of Doppler cooling in the case of strontium atoms \cite{Chalony:2011}.
\\~\\

\subsection*{Production of a $^{162}$~Dy BEC}

We now discuss the  production of a Bose-Einstein condensate after evaporative cooling in a crossed dipole trap (see Fig.~\ref{Fig4}). We used the $^{162}$Dy isotope as it exhibits a larger background scattering length, which enhances the elastic collision rate compared to the $^{164}$Dy isotope. We checked that the electric polarizability and the Doppler cooling work equivalently for the two isotopes, as expected since the nuclear spin is zero in both cases.

The optimization of the Doppler cooling stage allows us to reach a phase-space density of $\varpi = 5.7 (10) \times 10^{-4}$, and to load approximately $9\times 10 ^5$ atoms in a crossed dipole trap formed of the laser discussed in previous sections (ODT 1), a circular Gaussian beam with waist of $25$~\micro m operating at 1064 nm with a maximum output power of 45 W (ODT 2) and an elliptical Gaussian beam with waists of $63$~\micro m and $41$~\micro m operating at 1064 nm and with 9 W maximum output power (ODT 3). The circular Gaussian beam (ODT 2) is spatially modulated (at a frequency of $50$~kHz) through the use of a deflector which makes it effectively elliptic. The modulation is reduced through the evaporation in order to increase the collisional rate and maximize the evaporation efficiency. All three optical beams lie on the horizontal plane and form angles with respect to ODT 1 of -56$^\circ$ (ODT 2) and 30.6$^\circ$ (ODT 3). The magnetic field is kept at a fixed value of $1.45$~G, away from any Feshbach resonance.

A schematic representation of the evaporation procedure is shown in Fig.~\ref{Fig4} a). The evaporation efficiency given by $\gamma = - \text{d}\log{\varpi}/ \text{d}\log{N}$  is, for most of the evaporation protocol, close to 4 (see Fig.~\ref{Fig4}b ). Bose--Einstein condensation is then reached at a critical temperature of $120(20) $~nK. After further evaporative cooling, we obtain a quasi-pure BEC with $\sim 3 \times 10^4$ atoms in an harmonic trap with aspect ratio $\omega_z /\sqrt{\omega_x \omega_y} = 1.7$.
\\~\\
In conclusion, we have observed the tunability of the differential light-shift for the $626$~nm transition in the case of a thermal Dy cloud confined in a far-detuned, $1070$~nm, optical dipole trap. We observe that, for a given trapping beam polarization angle, a total cancellation of the differential light-shift can be achieved. This observation is in qualitative agreement with the most recent theoretical models as discussed in Section.~\ref{sec:theory} and provides valuable information on the excited state polarizability.
We demonstrate the relevance of the magic-polarization behaviour by optimizing a Doppler cooling stage which led us to reach a degenerate dipolar gas.
Furthermore, the magic-polarization behaviour opens the prospect of sideband cooling in optical lattices for the purpose of single site imaging \cite{Ott:2016}. 
\\~\\
This work is supported by PSL University (MAFAG project), European Union (ERC UQUAM \& TOPODY) and DIM Nano-K under the project InterDy. We thank Davide Dreon and Leonid Sidorenkov for contributions in earlier stages of the experiment. 
\bibliography{RLopes_BIB.bib}


\end{document}